\documentclass[astra]{copernicus}
\pdfoutput=1

\usepackage{amsmath}

\begin{document}

\title{Directional correlations between UHECRs and neutrinos observed with IceCube}

\author[2]{Robert Lauer$^1$ for the IceCube Collaboration}

\affil[1]{DESY, Platanenallee 6, D-15738 Zeuthen, Germany}
\affil[2]{http://icecube.wisc.edu}


\runningtitle{Correlations of UHECRs and IceCube neutrinos}

\runningauthor{R. Lauer}

\correspondence{R. Lauer, robert.lauer@desy.de}

\received{}
\pubdiscuss{} 
\revised{}
\accepted{}
\published{}


\firstpage{1}

\maketitle

\begin{abstract}

Ultra-high energy cosmic rays (UHECR) above an energy threshold of tens of EeV might undergo only small deflections due to interstellar magnetic fields.
Their arrival directions would then point to regions of possible hadronic acceleration processes, which are likely to be also sources of high energy neutrinos. 
To search for such cosmic accelerators, we present here the first high statistics analysis of directional correlations between neutrino candidates from the IceCube Observatory and UHECR events.
Data taken with IceCube in a configuration of 22 strings provided the basis for using published events from both the Pierre Auger Observatory and the HiRes experiment as reference directions in a search for coincidences with neutrinos.
The analysis was optimized according to strict blindness criteria and showed an excess of neutrino candidates close to UHECR directions with a probability of 1\% to occur as a random fluctuation, consistent with a background-only hypothesis.
An extension of this analysis to include newer IceCube data, taken with 40 strings and using a likelihood analysis, is discussed in the outlook.
\end{abstract}


\introduction

The successful operation of the HiRes experiment until 2004 and the growing data sample provided by the Pierre Auger Observatory (PAO) made high statistics studies of UHECRs possible. Nevertheless, their sources remain unknown and the composition features have not yet been conclusively determined~\citep{PaoComp,HiresComp}. The limited knowledge about astrophysical magnetic fields prohibits detailed predictions about their propagation, but at energies above $10^{19}$ eV, at least protons and light nuclei should undergo only small deflections on the order of a few degrees. Consequently, these cosmic rays will point back to regions of the sky where astrophysical sources achieve hadronic acceleration up to the highest energies, and are thus promising places to search for high energy neutrino emissions.

Up to now, no cosmic point sources of neutrinos at TeV or higher energies have been identified. Searches for these neutrinos are performed on data samples dominated by atmospheric background events and in consequence their discovery potential can be boosted by limiting the trials via testing only directions defined by UHECR events.
Such a multi-messenger approach has the primary objective to search for efficient neutrino emission connected with hadronic accelerators but could ultimately help to learn more about the distribution of UHECR sources and astrophysical magnetic fields.

\section{Instruments and data samples}

The IceCube Observatory~\citep{ICkarg} is currently the largest neutrino telescope and was used in this UHECR correlation search. Located in the deep ice at the geographic south pole, photomultipliers in digital optical modules~\citep{ICDAQ} are arranged on strings, carrying 60 modules each and reaching down to 2450 m below the surface. The array serves to measure Cherenkov light from charged particles, in particular secondaries from neutrino interactions, and a nanosecond resolution of photon arrival times allows reconstructing the direction of muon tracks. The deployment of a total of 86 strings in a volume of a cubic kilometer will be finalized until the beginning of 2011. 

The first UHECR correlation study has been performed with IceCube data collected over 276 days in 2007/2008 with a configuration of 22 strings. It was based on the muon track sample selected for the first point source search that included not only upward going muon tracks, but also events from above the horizon, a region previously excluded due to increased background fluxes~\citep{IC22uheps}.
The events from the northern sky are mainly induced by atmospheric neutrinos, while the southern sky is dominated by downward going atmospheric air shower muons.
The neutrino energy range covered with this selection for a hypothetical $E^{-2}$ signal spectrum reaches from TeV to PeV for upward going events, limited by absorption in the Earth. The energy range for downward going tracks covers PeV to EeV, because lower energies are suppressed through background reduction cuts.

\begin{figure}
\centering
	\includegraphics[width=230pt]{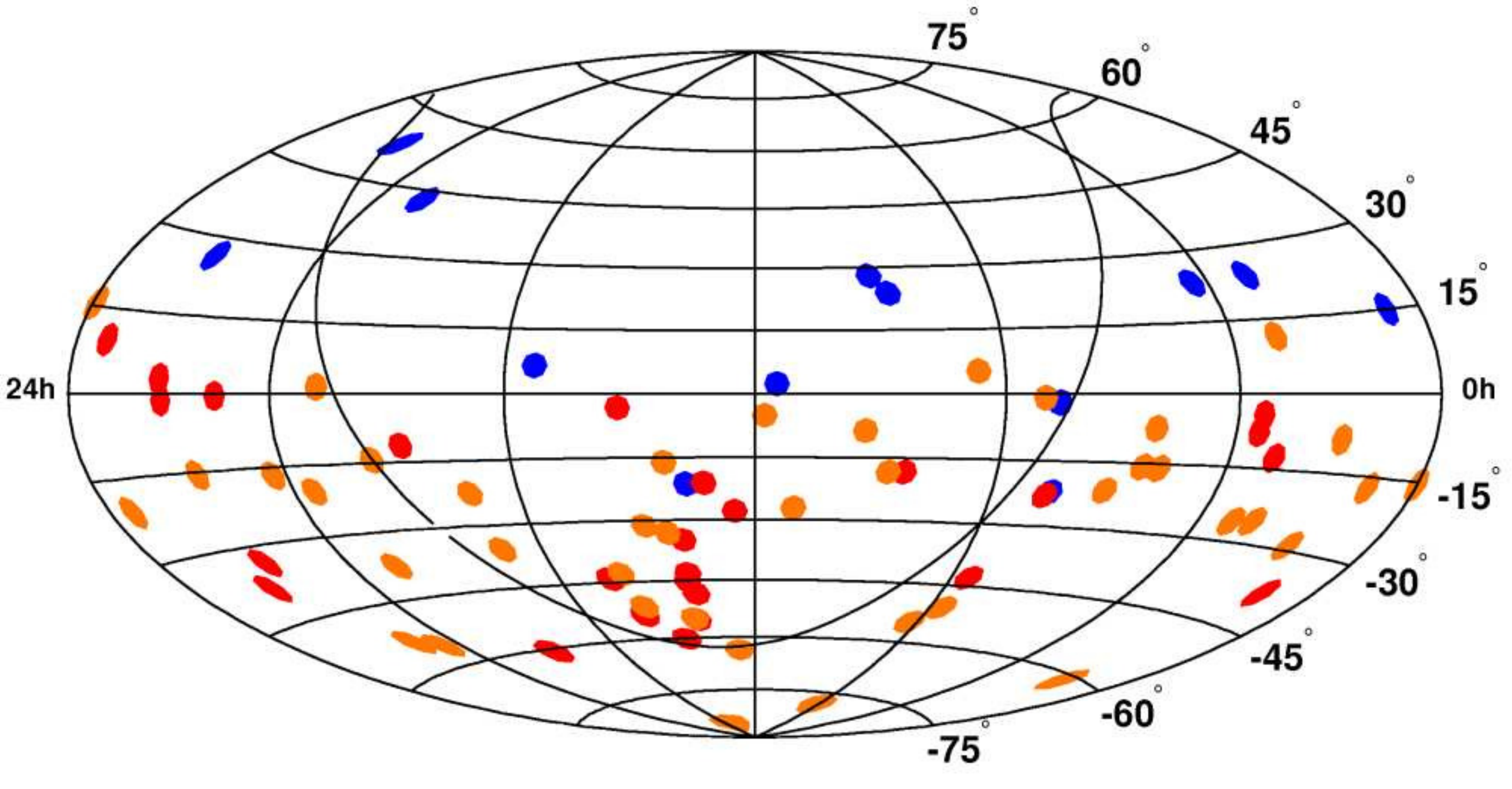} 
\caption[UHECR event directions.]{\label{fig_uhecrmap}Arrival directions of UHECR in equatorial coordinates, with the galactic plane shown as a curved line. 13 events are from HiRes (blue), 27 from the PAO 2007 sample (red) and 42 additional events (orange) released in 2010.}
\end{figure}

For these correlation searches, publicly available UHECR data from two extensive air shower detectors are used as reference directions.
One is the HiRes experiment, operating between 1999 and 2004 in stereo mode with two fluorescence detectors~\citep{HiResStereo2009}.
The other is the southern PAO array which is taking data since 2004 with a hybrid technique of fluorescence telescopes and Cherenkov detectors.
The first data on arrival directions of UHECRs published by PAO in~\citep{AugerCorr08} comprised 27 events, collected between January 2004 and August 2007.
The chosen energy threshold of 56 EeV was optimized in a search for correlations with nearby AGN
together with two other search parameters. 
In that analysis, the hypothesis of an isotropic distribution of UHECRs was rejected with a probability of 99\% and a correlation with the AGN sample was claimed. 
Recently, the results were updated in~\citep{AugerCorr10} with 42 additional events detected through the end of 2009. 
The degree of correlation with the AGN catalog has decreased for the full sample, but this has no impact on the neutrino analysis presented here, since only the reported coordinates of UHECR arrival directions themselves are considered. 
Following the PAO reports, the HiRes collaboration applied the same energy threshold as in the PAO study to all their events collected in stereo mode.
They found 13 events and published the arrival directions in~\citep{HiResCorr2008}.
HiRes could neither confirm a correlation with the AGN catalog nor find an indication of anisotropy in a separate test with a different event selection.

At the time of our first correlation search, only the original 27 PAO and 13 HiRes events had been made public. 
Because of the low signal expectations, we did not use IceCube 22 strings data with declinations below $-50^{\circ}$, leaving 35 UHECR arrival directions.
The new IceCube event selection for 40 strings covers all declinations
and thus can be correlated with the full sample of 82 UHECR events shown in Fig.~\ref{fig_uhecrmap}.

\section{Binned correlation method}

A possible correlation between UHECR directions and the IceCube data can be tested by counting neutrino candidates in circular bins of a pre-defined radius centred on the arrival directions of the charged cosmic rays. 
In~\citep{Petrovic2009}, an application of such an approach to generic neutrino and UHECR data samples is discussed.
To calculate the significance of the observation, the sum of events over all bins has to be compared to the expectation from pure background samples. 
These can be produced by randomizing the individual right ascension values of measured data, because the acceptance varies only with declination due to IceCube's location and the integrated live time of a year. 
Counting IceCube events over all bins is performed by testing if the minimum angular distance to any UHECR direction is smaller than the fixed bin radius to avoid double counting in overlapping bins. 
The distribution of results from scrambled samples is then integrated over all event numbers larger than the one observed. This yields the p-value, defined as the probability for the measured outcome to appear in a background-only scenario. 

Simulated sky maps with added signals are required to assess the discovery potential and optimize the analysis for realistic expectations.
Since only the UHECR directions can be used as reference positions for simulated neutrino sources, the effects of magnetic fields on the ionized nuclei have to be taken into account.
There are few established facts about astrophysical magnetic fields and predictions vary concerning their strengths and orientations. 
The PAO collaboration assumes deviations around $3^{\circ}$~and up to $7^{\circ}$ in~\citep{AugerCorr08} and references therein. Other authors like~\citep{Kachelriess2005} also predict average deflections to be on the order of a few degrees in the energy range under consideration, though possibly stronger deviations occur close to the galactic center. Extensive simulations for galaxy clusters discussed in \citep{Dolag2005} show that the impact of extra-galactic fields on UHECR propagation is negligible.
These predictions are only valid for proton UHECRs, and deviations for heavier nuclei are expected to be significantly larger which would then strongly limit the prospects of a neutrino correlation approach.

To account for unknown magnetic deflections in the simulation, neutrino sources were individually shifted with respect to the UHECR directions. Based on the above references, a width of $3^{\circ}$~was chosen for a Gaussian function according to which randomly orientated angular displacement values were generated in each trial individually for all UHECR directions. The angular resolution of the air shower reconstruction for the chosen data is better than $0.9^{\circ}$~\citep{HiResCorr2008, AugerCorr10} and was thus considered as negligible compared to the deflection uncertainty.

In the absence of any information on individual muon neutrino expectations for the hypothetical sources, equal point source fluxes from all UHECR directions, following an $E^{-2}$ spectrum, were assumed in the simulation.
Accounting for the variations of IceCube sensitivity with declination as shown in~\citep{IC22uheps}, average numbers of neutrinos were derived for each source direction and used to generate event numbers in each trial according to Poisson distributions. 
Simulated signals were placed into scrambled sky maps by distributing them around the shifted source positions with angular deviations that followed the IceCube point spread function.

The distributions of binned event counts for simulations with and without signal can be well fitted with Gaussian functions. They allow deriving a $5 \sigma$ (one-sided Gaussian tail) threshold from the background scenario and then determining the flux strength for which the probability of such an excess reaches 50\%, called the discovery potential.
The bin radius is the only free parameter in this correlation approach.
For 22 strings data, various radii were tested with respect to the best discovery potential for simulated signals and an optimal value of $3^{\circ}$ was found. Simulations with an increased width for the source shifts showed that the optimal value does not become significantly larger even for greater deflections due to the additional background covered by the bins. Details on the optimization are provided in~\citep{Lauer2010}.
Since the prescription was fixed before applying it to the true event coordinates, no trial factor enters into the final p-value.

The resulting discovery potential for the 22 strings search, expressed as an $E^{-2}$ muon neutrino flux normalization for $d\Phi/dE=\Phi_0 E^{-2}$, is $\Phi_0 =1.0\cdot10^{-8}$ GeV cm$^{-2}$ s$^{-1}$. This is approximately a factor of three smaller than the upper limit on the average muon neutrino point source flux for 35 equal sources derived from the best diffuse neutrino limit in the same energy range in~\citep{AmandaDiffuse08}.

\section{Results from the search on IceCube 22 strings data}

After applying the analysis prescription to the IceCube data by scanning the 35 bins of $3^{\circ}$~radius, a total of 60 events were observed. The mean background expectation based on scrambled sky maps was 43.7 events.
The probability for an excess of the same or a larger magnitude within the background-only scenario is 0.0098, equivalent to~$2.3\sigma$.
This result is compatible with a background fluctuation.
No particularly strong clustering of IceCube events on smaller scales than the bin size was observed and the muon energy estimates for the 60 events featured no unusual deviations from the full distribution of background data.

Under the benchmark emission hypothesis of equal neutrino fluxes from all 35 PAO and \textsc{HiRes} UHECR directions, the result can be used to derive an upper limit on the muon neutrino flux normalization per source, yielding
$\Phi_{\mbox{\scriptsize lim}}=0.9 \cdot10^{-8}$ GeV cm$^{-2}$ s$^{-1}$ for a mean magnetic deflection of $3^{\circ}$. 
Since this uniform emission scenario is only a convenient framework for interpreting the event counts, the limit should be considered at best as an average level of probed flux strengths.

\section{Outlook for the search with IceCube 40 strings data}

The newly released PAO arrival directions from~\citep{AugerCorr10} allow us to extend the analysis to a total of 82 UHECR reference directions.
Based on this a search on the independent IceCube data collected with 40 strings over 376 days is in progress. 
The event selection is adopted from the all-sky point source search as presented in~\citep{ic40ps}, comprising 36,900 muon tracks consistent with expectations from atmospheric background simulations.
Scrambled sky maps and simulations with added neutrino signals from shifted sources were produced in the way described above to study the performance of the binned search as well as a new unbinned likelihood approach.

The differences in the point source event selection for 40 strings compared to the 22 strings sample made it necessary to apply an additional cut on the energy estimator to perform the binned search in a similar way.
The cut was optimized for sensitivity and is parametrized as a zenith dependent threshold, decreasing from the downward to the upward going regime. It resulted in 6013 remaining neutrino candidate events.
Applying the binned search method to the simulations proved that the change of the discovery potential with bin radius is again small. Hence it was decided to keep the value of $3^{\circ}$ for consistency with 22 strings.
The resulting discovery flux threshold for the binned search is $\Phi_{\mbox{\scriptsize lim}}=0.7 \cdot10^{-8}$ GeV cm$^{-2}$ s$^{-1}$. The impact of the additional UHECR directions is small, because most of them lie in the southern hemisphere where the $E^{-2}$ acceptance is limited due to the high energy threshold.

To improve the sensitivity by including individual event parameters in the probability calculation, we adopted the unbinned likelihood method from~\citep{BraunUBL} to the correlation search. Its use in IceCube point source searches has been established in~\citep{IC22ps}. 
The likelihood formula Eq.~(\ref{ubllh}) is a product over the contributions with index $i$ from all IceCube events. As done in~\citep{ic40ps} for stacking searches on astronomical object classes, we make the signal term a sum over the contributions from many reference points, in our case the 82 UHECR arrival directions $\mathbf{u}_j$. 
The expectation for the average number of neutrinos which would be detected from direction $j$ for a common source spectrum $E^{-\gamma}$ is included via the acceptance weighting function $R(j,\gamma)$.
It depends on the position of each hypothetical source and corrects for the declination-dependent energy response of the detector. 
The sum over the UHECR events with index $k$ serves as normalization. 
\begin{align}
\label{ubllh}
\mathcal{L}  (n_s, \gamma) = \prod_i (& \frac{n_s}{n_{\mathrm{tot}}} \sum_{j} R(j,\gamma)
S_j(\mathbf{x}_i,\sigma_i,E_i,\gamma)/\sum_{k} R(k,\gamma) \notag \\
& + \left(1- \frac{n_s}{n_{\mathrm{tot}}}\right) B(\mathbf{x}_i,E_i) \;) \;\;. 
\end{align}
By minimizing the negative logarithm of $\mathcal{L}$, the best fit values for the total number of signal events $n_s$ and a common spectral index $\gamma$ is determined. The comparison of the likelihood value with that for $n_s=0$ yields the p-value for the null hypothesis.
The background probability density function (PDF) $B(\mathbf{x}_i,E_i)$ depends on the individual event energy estimator $E_i$ and the declination of its location $x_i$ and is derived directly from data. For the PDF  $S_j(\mathbf{x}_i,\sigma_i,E_i,\gamma)$ representing the signal hypothesis with respect to direction $j$, the energy dependence is obtained from neutrino simulations for spectral indices $\gamma$ between 1 and 4. In the spatial part, the point spread function is represented by a Gaussian function with a width $\sigma_i$ that is obtained individually as a direct reconstruction error estimate~\citep{ic40ps}. This function has to be convoluted with the probability distribution for a source shift due to magnetic deflection, for which we use again a Gaussian with a width of $\sigma_s = 3^{\circ}$, as discussed with respect to the simulation above. This convolution results in a broadened Gaussian function for the spatial PDF,
\begin{equation}
 S_{j,spatial}(\mathbf{x}_i,\sigma_i) = \frac{1}{2\pi (\sigma_{i}^2+\sigma_{s}^2)} \;\exp \left(-\frac{|\mathbf{x}_i-\mathbf{u}_j|^2}{2 (\sigma_{i}^2+\sigma_{s}^2)}\right)\;.
\end{equation}

The algorithm is applied to sky maps of the full 40 strings sample, with simulated neutrinos injected from shifted sources as outlined above. The resulting discovery potential for the scenario of equal $E^{-2}$ fluxes from all 82 directions is $\Phi_{\mbox{\scriptsize lim}}=0.2 \cdot10^{-8}$ GeV cm$^{-2}$ s$^{-1}$. This is equal to an average of 52 signal neutrinos in the point source sample of 36,900 events. The improvement over the binned method is largely achieved through the direct use of event energy estimators and individual reconstruction errors and the corresponding PDFs.
The acceptance function $R$ gives additional discrimination power by weighting directions according to their hypothetical contributions, but this will only be a major advantage if the assumption of many more or less equal sources is justified.

Testing the same algorithm on simulations of neutrino sources with the width of the source shift distribution increased to $6^{\circ}$ raises the discovery flux threshold by approximately 50\%. This reflects a good degree of flexibility, in particular when compared to the binned search where the increased shifts worsen the discovery potential by a factor of 2.

\conclusions
We performed the first large statistics analysis of directional correlations between 35 UHECR arrival directions reported by PAO and HiRes and IceCube data collected over one year with 22 strings. The observed excess has a 1\% chance to occur as a background fluctuation and is consistent with the background-only hypothesis. The IceCube data collected with 40 strings and newly released UHECR directions are currently being used to perform an extended search, employing a likelihood approach to improve the sensitivity compared to the binned event counting.

\bibliography{lauer_ecrs}
\bibliographystyle{copernicus}

\end{document}